\documentclass[10pt, reqno]{amsart}
\usepackage{float}
\usepackage{graphicx}
\usepackage{orcidlink}
\usepackage{multirow}
\usepackage{indentfirst,csquotes}
\usepackage{amssymb,amsthm,amsmath}
\usepackage{xcolor,paralist,hyperref,fancyhdr,etoolbox}
\usepackage{algorithm}
\usepackage{algorithmic}

\hypersetup{ colorlinks=true, linkcolor=black, filecolor=black, urlcolor=black, citecolor=black }

\makeatletter
\def\section{\@startsection{section}{1}%
  \z@{.7\linespacing\@plus\linespacing}{.5\linespacing}%
  {\normalfont\bfseries}}

\def\subsection{\@startsection{subsection}{2}%
  \z@{.5\linespacing\@plus.7\linespacing}{.3\linespacing}%
  {\normalfont\bfseries}}

\let\oldsection\section

\renewcommand{\section}{%
  \@ifstar{\section@star}{\section@nostar}%
}

\def\section@star#1{%
  \oldsection*{\MakeUppercase{#1}}%
}

\def\section@nostar#1{%
  \oldsection{\MakeUppercase{#1}}%
}

\renewcommand{\@seccntformat}[1]{%
  \csname the#1\endcsname\quad$|$\quad}
\makeatother

\begin{document}

\begin{center}
{\Large\bfseries Smooth Total variation Regularization for Interference\\ Detection and Elimination (STRIDE) for MRI}

\vspace{1em}

Alexander Mertens$^{1,2}$\,\orcidlink{0000-0001-5172-0444}, 
Diego Martinez$^{1,2}$\,\orcidlink{0009-0004-4272-7782},
Amgad Louka$^{1}$\,\orcidlink{0009-0003-5595-5381}, 
Ying Yang$^{1,2}$\,\orcidlink{0009-0002-8804-566X}, 
Chad Harris$^{4}$\,\orcidlink{0009-0001-2004-468X}, 
Ian Connell$^{1,2,3,5}$\,\orcidlink{0000-0002-7722-3603}

\vspace{0.5em}

{\small
$^1$Biomedical Engineering, University Health Network, Toronto, ON, CA\\
$^2$Medical Biophysics, Temerty Faculty of Medicine, University of Toronto, Toronto, ON, CA\\
$^3$CenteR for Advancing Neurotechnical Innovation to Applications, University Health Network, Toronto, ON, CA\\
$^4$Research and Development, Synaptive Medical, Toronto, ON, CA\\
$^5$The KITE Institute, University Health Network, Toronto, ON, CA\\
}

\vspace{0.5em}

{\small Correspondence: Ian.Connell@uhn.ca}

\vspace{0.5em}

{\small\today}
\end{center}

\vspace{1em}

\noindent{\small\textit{Funding:} This work was partially supported by Ontario Research Fund Research Excellence program.}

\vspace{0.5em}

\noindent{\small\textit{Keywords:} MRI, Electromagnetic Interference, STRIDE, EDITER, low-field MRI}

\begin{abstract}
MRI is increasingly desired to function near electronic devices that emit potentially dynamic electromagnetic interference (EMI). To accommodate for this, we propose the STRIDE method, which improves on previous external-sensor-based EMI removal methods by exploiting inherent MR image smoothness in its total variation. STRIDE measures data from both EMI detectors and primary MR imaging coils, transforms this data into the image domain, and for each column of the resulting image array, combines and subtracts data from the EMI detectors in a way that optimizes for total-variation smoothness. Performance was tested on phantom and in-vivo datasets with a 0.5T scanner. STRIDE resulted in visually better EMI removal, higher temporal SNR, larger EMI removal percentage, and lower RMSE than standard implementations. STRIDE is a robust technique that leverages inherent MR image properties to provide improved EMI removal performance over standard algorithms, particularly for time-varying noise sources.
\end{abstract}

\bigskip

\section{Introduction}\label{sec1}

\subsection{Motivation}

Magnetic resonance imaging (MRI) is a highly sensitive imaging modality that is unmatched in its ability to non-invasively visualize soft tissue. Due to its high sensitivity, MRI is also highly susceptible to electromagnetic interference (EMI). Any devices emitting EMI near the Larmor frequency contaminate the measured signal and severely degrade image quality. To prevent EMI contamination, MR scanners operate inside dedicated radiofrequency (RF) shielded rooms (Faraday cages), allowing them to be isolated from all outside electronics and potential sources of EMI.

However, many emerging applications of MRI require the introduction of foreign devices into its environment; for example, interventional MRI (iMRI) or acute imaging \cite{Iordachita, jellema2023, wach2025, Marincola2018MRI, Madewell2019Neurosurgical, rogers2021, pawlik2024, KUMAR2024107662, puhrwesterheide2022cost, orlando}. These application specific peripheral devices can severely degrade image quality if not carefully designed to produce minimal impact within the MR environment. Ensuring adequate MR compatibility of foreign devices is a daunting challenge, particularly in the operating room (OR), emergency department (ED), or intensive care unit (ICU) space, where specialized equipment exists that was never intended to be co-located with an MRI unit. The problem is exacerbated by novel, point-of-care, small magnetic footprint MRI systems, where electronic devices can be placed closer to the MR system without being damaged by the stray magnetic field, or becoming a safety risk \cite{kuoy2022poc_mri, panther2019dedicated}.

\subsection{Existing Methods}

Various methods have been developed to eliminate EMI in MRI. Passive solutions integrate local shielding into the scanner design but must have an opening to accommodate the patient \cite{OReilly2021, Cooley2021}. Even though the shield opening is small compared to the wavelength at the Larmor frequency for low-field MRI, the patient acts as an antenna, bringing noise into the bore unless also covered by shielding, which is impractical in many scenarios \cite{OReilly2021}. 

Progress in algorithmic EMI removal for MR has been made across many fields including geophysics, biomedical instrumentation, surface NMR, and MRI \cite{gamble1979, osti_7023612, spies1988, espy2015, walsh2008, muller-petke2020, ren2017, EDITER, letcher1989}. Rather than passive shielding, many of these methods use EMI sensors whose purpose is to measure EMI independently of the primary signal of interest. Early MRI adaptations relied on calibration data to characterize the relationship between primary imaging coils and EMI sensors, limiting their effectiveness when the EMI is dynamic.

In the field of MRI, External Dynamic Interference Estimation and Removal (EDITER) has become a widely used method for deterministic EMI removal \cite{EDITER}. EDITER clusters acquired data into groups where the EMI is similar within each group. Then, for each group, a linear transfer function that maps the EMI sensors to the primary imaging coils is estimated and used to clean the data. The EMI sensors used in the EDITER method consist of an EMI-only pick-up coil (similar to an MR receive coil) and an electrode. 

Clustering the data into stationary groups makes EDITER robust to changing EMI sources, and is therefore  suitable for uncontrolled environments. When the number of EMI sensors is small relative to the amount of imaging data in a group, the transfer function may underfit and not remove all of the EMI. Conversely, if the number of EMI sensors is high relative to the amount of imaging data in a group, the transfer function may overfit and remove image data as well as EMI \cite{EDITER}. An additional issue (which is common to all methods utilizing EMI sensors) is that in the process of removing the EMI the thermal noise of the sensors will be added to the imaging coils.

More recently, several machine learning (ML) techniques have been developed for EMI removal in MRI. These approaches typically use dedicated EMI characterization windows during each TR where no RF excitation occurs, allowing simultaneous EMI measurement by both primary receive coils and EMI sensing coils. Modern methods include convolutional neural networks that model the complex relationship between EMI sensors and receive coils \cite{ML1, ML2}, and more advanced architectures like Deep-DSP (Deep Learning Direct MR Signal Prediction) that directly predict EMI-free MR signals rather than subtracting estimated EMI \cite{ML1}. After training on EMI characterization data, these models can predict and remove EMI components from the actual MR signal acquisition window. 

In this work, we propose Smooth Total variation Regularization for Interference Detection and Elimination (STRIDE), which we compare against EDITER. STRIDE provides methods of improving EDITER and other deterministic transfer-function based EMI removal methods. Our contributions are twofold:

\begin{enumerate}
    \item The addition of a compressed-sensing (CS) inspired total variation (TV) heuristic to mitigate the overfitting problem mentioned above, allowing EMI removal with fewer noise coils and better accommodation for time varying noise sources.
    \item Demonstration that the performance of any noise coil method is affected by the SNR of the EMI sensors.
\end{enumerate}

Deterministic methods were selected over ML techniques for their transparency and interpretability, universal compatibility with all pulse sequences, and independence from data collection and training requirements.

\section{Theory}\label{theory}

\subsection{Total-Variation (TV) Regularization in EMI Removal}\label{TVReg}

Compressed sensing is widely used in reconstruction of undersampled MR data \cite{mertens2025, Feng2014, Feng2016, Feng2020, Weiss2019, Demerath2020, Hainc2019}. In general, CS produces an image that is both consistent with measured \textit{k-space} data in an $L2$ sense, and is sparse in some domain (i.e., a small L1 norm after a particular linear transformation). The prior knowledge that MR images are sparse in various domains allows the recovery of images even when the measured data are noisy and do not meet the Nyquist sampling rate \cite{lustig2008}. A typical CS problem formulation is as follows:

\begin{equation}
\hat{x} = \underset{x}{argmin} ||y -\mathrm{F}x||^2_2 + \lambda||\mathrm{W}x||_1
\label{eqn:CSproblem}
\end{equation}

Here, $y$ is measured \textit{k-space}, $\mathrm{F}$ is a Fourier operator that may include coil sensitivity profiles, $x$ is a candidate for the final reconstructed image, $\mathrm{W}$ is a sparsifying transform, $\lambda$ determines the tradeoff between data consistency and sparsity, and $\hat{x}$ is the final estimated image.

Although solving for the transfer function from EMI sensor to imaging coil is not a CS problem, we hypothesize that leveraging the knowledge that MR images exhibit low L1 norms under sparsifying transforms can improve the performance of EMI removal algorithms. First, we start with a standard EMI removal problem formulation:

\begin{equation}
\begin{split}
\mathrm{\hat{A}} &= \underset{\mathrm{A}}{argmin} ||y -\mathrm{UA}||^2_2\\
\hat{x} &= y - \mathrm{U\hat{A}}
\end{split}
\label{eqn:standardEMIRemoval}
\end{equation}

Here, $y$ is measured data in one image coil, $\mathrm{U}$ is the subspace that describes the noise subspace, $\mathrm{A}$ is a candidate for the optimal noise subspace coefficients, $\hat{\mathrm{A}}$ are the optimal noise subspace coefficients, and $\hat{x}$ is the estimated EMI-free data. In the case of EDITER, this problem is solved once per temporal group with $\mathrm{U}$ determined by EDITER variables $\Delta kx$ and $\Delta ky$ and the data in the EMI sensors. This problem formulation assumes no correlation between measured data $y$ and noise subspace $\mathrm{U}$. In practice, small correlations may exist. The effect is that when the temporal groups are small (i.e the columns of $\mathrm{U}$ are numerous relative to the length of vector $y$), overfitting may occur and image data may be removed when solving this equation. This worsening of image quality as temporal group sizes decrease was noted by the authors of EDITER \cite{EDITER}, and the solution was to keep the group sizes relatively large (a small number of temporal groups) to reduce overfitting.

While this solution is effective, it may fall short in cases where small temporal groups are needed (as could be the case with non-stationary noise). To address this, we propose that incorporating L1 norm regularization in a sparse domain, as in CS, allows for reconstruction using a $y$ vector in the image domain that is short in length, similar to numerous small temporal groups as in EDITER (e.g., one PE line per temporal group). This improves overall robustness since it does not assume zero correlation between noise subspace and measured data, and does so without degradation of image quality. We propose the following problem formulation in the image domain: 

\begin{equation}
\begin{split}
\mathrm{\hat{A}} &= \underset{\mathrm{A}}{argmin} ||\mathrm{W}\left( y_{img} -\mathrm{UA} \right)||_1\\
\hat{x}_{img} &= y_{img} - \mathrm{U\hat{A}}
\end{split}
\label{eqn:STRIDERemovalL1}
\end{equation}

Here, $y_{img}$ is one column of a $kx \times ky$ coil image, $\mathrm{W} \in \mathbb{C}^{kx-1 \times kx}$ is a total variation matrix, $\mathrm{U} \in \mathbb{C}^{kx \times \mathrm{N_c} \Delta y}$ is the noise subspace determined by variable $\Delta y$, and the corresponding columns in the $kx \times ky$ EMI sensor images (after Fourier transform to the image domain), $\mathrm{A}\in \mathbb{C}^{\mathrm{N_c} \Delta y }$ are the candidate coefficients for the noise subspace, $\hat{\mathrm{A}} \in \mathbb{C}^{\mathrm{N_c} \Delta y}$ are the optimal noise subspace coefficients, and $\hat{x}_{img}$ is the corresponding estimated column of the EMI-free image. Similar to $\Delta ky$ and $\Delta kx$ in EDITER, $\Delta y$ and $\Delta x$ determine which adjacent columns of the EMI sensor data are used in building the noise subspace. Since the operation is done in the image domain, variables $\Delta y$ and $\Delta x$ are used instead of $\Delta ky$ and $\Delta kx$.

This problem formulation ensures that EMI is subtracted from the coil images such that the resulting image has a sparse total variation in the readout direction.

However, this L1 norm minimization does not have a closed form solution and gradient descent methods would typically be used to solve it. Since one problem of this form would have to be solved for each column of the EMI-contaminated image, leading to a large computation time, we relax the problem to be an L2 norm minimization problem:

\begin{equation}
\begin{split}
\mathrm{\hat{A}} &= \underset{\mathrm{A}}{argmin} ||\mathrm{W}\left( y_{img} -\mathrm{UA} \right)||_2^2\\
\hat{x}_{img} &= y_{img} - \mathrm{U\hat{A}}
\end{split}
\label{eqn:STRIDERemovalL2}
\end{equation}

This gives in a closed form solution for $\hat{\mathrm{A}}$:
\begin{equation}
\hat{\mathrm{A}} = \left( \mathrm{U^H W^H WU}\right)^{-1} \mathrm{U^H W^H W} y_{img}
\label{eqn:ASol}
\end{equation}

Although this relaxed problem promotes smoothness over sparsity, we hypothesize that it still results in an improvement over minimizing MSE as in the standard EMI removal problem.

Pseudocode for the algorithm is provided in algorithm \ref{alg:algo} below. In theory, this algorithm is independent of sampling pattern since it operates on EMI sensor and imaging coil data after an inverse uniform or non-uniform Fourier transform has been applied to each coil.

\begin{algorithm}[H]
\caption{Pseudocode for STRIDE}\label{alg:algo}
\begin{algorithmic}
    \STATE Inverse Fourier transform data from every coil into the image domain
    \FOR{$i \in \{0, 1, \ldots, n-2\}$}
        \STATE $\mathrm{W}_{i,i} = -1$
        \STATE $\mathrm{W}_{i,i+1} =1$
    \ENDFOR
    \FORALL{image columns $y_{img} \in \mathbb{C}^{kx}$}
    \STATE Construct $\mathrm{U} \in \mathbb{C}^{kx \times \mathrm{N_c} \Delta y}$ from the EMI sensors
    \STATE Compute coefficients: $\hat{\mathrm{A}} = (\mathrm{WU})^{\dagger} \mathrm{W} y_{img}$
    \STATE Remove EMI: $\hat{x}_{img} = y_{img} - \mathrm{U}\hat{\mathrm{A}}$
\ENDFOR
\end{algorithmic}
\end{algorithm}

\subsection{Importance of EMI sensor SNR}\label{EMICoilSNR}

He et al, demonstrated the relationship between an external ring sensor diameter and its ability to remove EMI \cite{He2022Active}. This suggests that the SNR of the EMI sensor influences performance. To demonstrate this, consider the simple example of one imaging coil and one noise coil. They can be described as follows:

\begin{equation}
\begin{split}
y_{noisy} &= y + w_{img}, \quad w_{img} \sim \mathcal{N}(0, \sigma_{img}^2)\\
x_{noisy} &= x + w_{emi}, \quad w_{emi} \sim \mathcal{N}(0, \sigma_{emi}^2)
\end{split}
\label{eqn:NoisyMeasurements}
\end{equation}

Here, $w_{img}$ is the measurement noise of the imaging coil, and $w_{emi}$ is the measurement noise of the noise coil. The EMI free image can be calculated as follows:

\begin{equation}
\hat{y}_{noisy} = y + w_{img} + ax_{noisy} + aw_{emi}
\label{eqn:NoisyTransfer}
\end{equation}

Here, the noise variance in the EMI-free image is proportional to the noise variance in the EMI sensor, showing the dependence of the EMI-corrected image SNR on the EMI detector SNR.

\section{Methods}

All experiments were performed on a head-specific 0.5T scanner (Synaptive Medical, Toronto, Canada) equipped with a 16 channel head-coil and 2 custom-built EMI sensors (Figure \ref{sup:figure_s1}).

\subsection{EMI sensor Construction}

Two in-house EMI detection coils were produced, each consisting of a 16 AWG copper wire 2 loop solenoid with a 25 mm diameter. These EMI detection coils, each connected to a preamplifier [Synaptive Medical; Toronto, Canada], were placed in an orthogonal configuration to cover the XY plane, behind the head coil in the bore of the system.

\subsection{Data Acquisition}

Two phantom datasets were acquired on a small ACR phantom using a 2D gradient echo (GRE) pulse sequence with a flip angle of 90 degrees, a matrix size of 256x256, 4-mm slice thickness, a TR of 50ms, and a TE of 7.238ms. The first was collected on a single contrast disc slice with an FOV of 25.6 cm, and the second was collected on the resolution dots slice with an FOV of 15 cm. These two slices were chosen to assess the effects on contrast and resolution of each reconstruction technique. For each slice, 64 images with one average each were acquired each under five EMI scenarios: (1) baseline acquisition (no EMI), (2, Square) function generator operating at 21.28 MHz with amplitude modulation by a 10 kHz square wave at -50 dB, (3, White noise) function generator at 21.28 MHz with amplitude modulation by white noise at -50 dB, (4, Sweep) function generator at 21.28 MHz with frequency sweeping ±10 kHz at a rate of 1 Hz to test the algorithms in the presence of non-stationary EMI, and (5, Vital sign monitor) Welch Allyn (WA) vital sign monitor with pulse oximeter and thermometer placed near the scanner bore. 

In-vivo brain data was collected on two human volunteers using the contrast disc phantom protocol. Volunteer imaging was conducted under informed consent and with University Health Network Research Ethics Board approval (REB\#23-5304). Similar to the phantom datasets, 64 single average images were collected under the same five EMI scenarios, the only difference being that the vital sign monitor was connected to the subject rather than being placed near the bore of the scanner.

Additionally, a single 15 average brain image was acquired under EMI scenario (2) using a GRE 2D pulse sequence with a flip angle of 30 degrees, a matrix size of 256x512, a 5-mm slice thickness, an FOV of 30-cm, a TR of 12.852-ms, and a TE of 6.798-ms. The purpose of this dataset was to provide the ideal conditions for each algorithm to eliminate EMI (i.e., low rank noise source to enable denoising of the EMI sensor, many averages for implicit denoising, and a flip angle closer to the Ernst angle).

\subsection{Data Reconstruction}

Each of the sets of 64 images were reconstructed coil-by-coil in the following ways: STRIDE with $\Delta y = 7$, and no EMI sensor denoising, EDITER with each phase encode line as a temporal group, similar to Huang et al. \cite{HUANG201852}, $\Delta ky=7$, $\Delta kx = 1$, and no EMI sensor cleaning (referred to as EDITER A), and EDITER with the number of clusters determined by the k-means algorithm described in their paper, $\Delta ky=1$, $\Delta kx = 1$, and no EMI sensor cleaning (referred to as EDITER B). For the STRIDE reconstructions, $\Delta y$ refers to the number of adjacent columns from the EMI sensors in image space used to build the noise subspace, similar to how $\Delta ky$ is the number of adjacent temporal groups used to make the noise subspace in EDITER. These parameters were chosen for STRIDE and EDITER B, the implementation as outlined in the original paper, since they provided the best reconstruction results. The parameters for EDITER A were chosen to match the parameters of STRIDE to show the advantage of regularizing for total variation rather than MSE. A pre-whitening matrix was computed using a short noise-only scan prior to each image acquisition. Then, channel combination was done using a standard sum-of-squares (SoS) method.

To demonstrate the effect of EMI sensor SNR, single average images from EMI scenario (2) were reconstructed with EMI sensor cleaning adhering to Gavish et al \cite{gavish2014optimalhardthresholdsingular}.

Channel images for the 15 average acquisition were reconstructed on an average-by-average basis with EMI sensor cleaning. EMI removal was performed using STRIDE and EDITER B. Channel combination and pre-whitening was performed using the vendor’s stock algorithm and combined images were complex averaged together \cite{gavish2014optimalhardthresholdsingular}. Lastly, the vendor’s image unwarping and shading corrected were applied.

\subsection{Evaluation Metrics}

Three metrics were computed on the phantom datasets to quantitatively evaluate their performance: SNR (SNR), EMI removal percentage, and RMS Error (RMSE)\cite{EDITER}.

The SNR was determined by taking the mean divided by the standard deviation of each voxel across the 64 acquisitions.

EMI removal percentage was computed voxelwise by taking the percent change of the standard deviation of the corrected images with pre-whitening to the standard deviation of the EMI corrupted images without pre-whitening.

RMSE was computed per voxel across the 64 images, using the mean of the baseline acquisitions as ground truth.

A binary mask made from baseline was used for each phantom dataset to separate the phantom from the background. On voxels containing phantom information, the mean SNR and EMI removal percentage was computed for each EMI removal method, as well as p-values from a Welch's t-test to determine any statistical significance in the difference between the EMI removal methods.

Lastly, to demonstrate the effect of denoising the EMI detectors, the standard deviation of a background region was computed in three cases: a square EMI corrupted \textit{in-vivo} brain scan, the EMI corrected image using STRIDE without denoising the EMI detectors, and the EMI corrected image using STRIDE with denoising the EMI detectors. Square wave noise was used to demonstrate this since it is low rank, allowing the EMI detectors to be denoised effectively.

\section{Results}

\begin{figure}[H]
  \centering
  \includegraphics[width=0.85\textwidth]{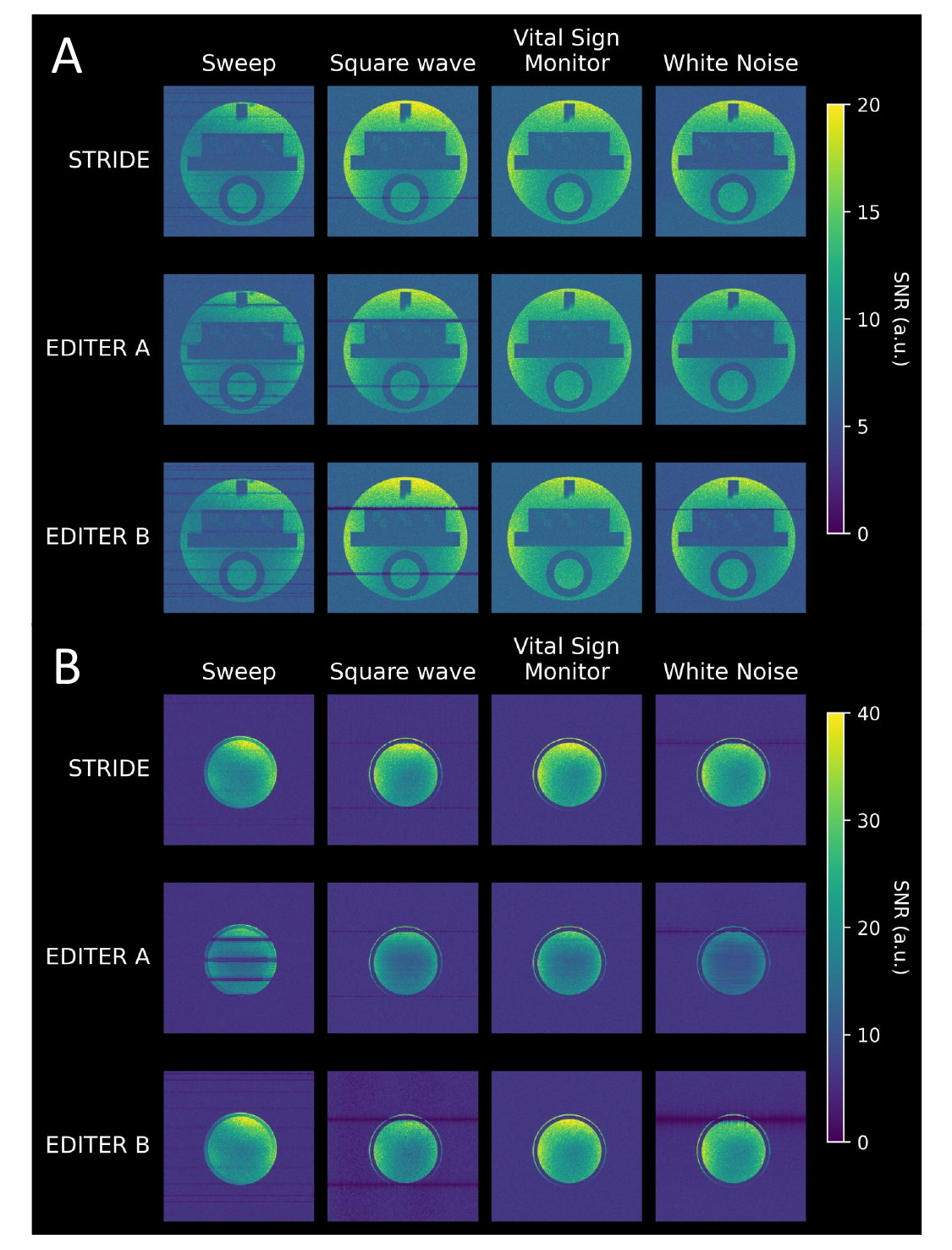}
  \caption{Visual comparison of voxel-wise SNR for each EMI removal method for \textbf{(A)} resolution dots and \textbf{(B)} contrast discs in a standard ACR phantom. A higher SNR is better. STRIDE is the proposed method, whereas EDITER A and B are two implementations of the standard EMI removal method \cite{EDITER}. EDITER A uses one phase encode line per temporal group, while EDITER B dynamically adjusts the number of phase encode lines per temporal group as outlined in the EDITER publication \cite{EDITER}. The resolution dots in \textbf{(A)} show overall similar SNR between methods, with a small dark spot in the upper-middle third section for STRIDE. In areas corrupted by narrow-band EMI, STRIDE shows consistently higher SNR, indicating superior EMI removal in those regions. The contrast discs in \textbf{(B)} show overall superior SNR in STRIDE compared to EDITER A and EDITER B, again with particular superior performance in areas of narrow-band EMI corruption.}
  \label{fig:figure1}
\end{figure}

\begin{figure}[H]
  \centering
  \includegraphics[width=\textwidth]{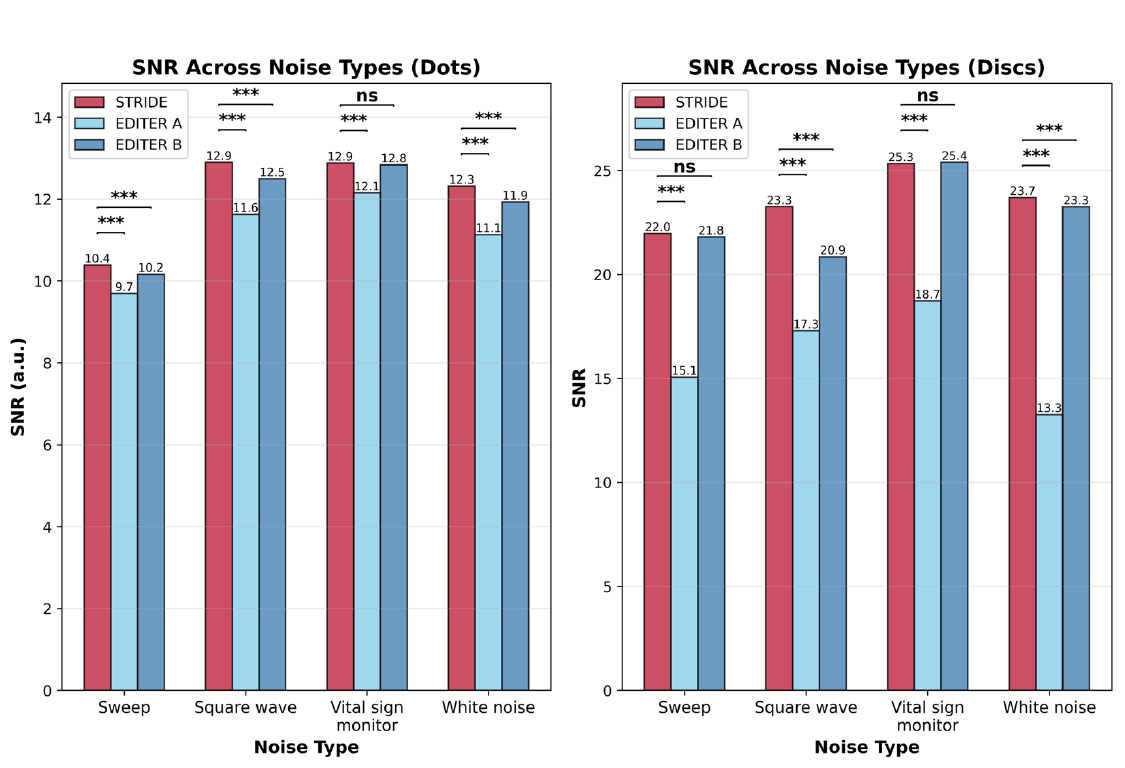}
  \caption{Graphical comparison of the mean SNR across non-background voxels in Figure \ref{fig:figure1} for each EMI removal method for\textbf{(left)} the resolution dots and \textbf{(right)} contrast discs in a standard ACR phantom, as well as the statistical significance of the difference in means as determined by a welch's t-test. $*$ indicates $ p < 0.05$, $**$ indicates $p < 0.01$, and $***$ indicates $p < 0.001$, and ns indicates no significance. A higher mean SNR is better. Overall, there were more cases for which STRIDE significantly outperformed both EDITER A and EDITER B, indicating a statistically significant improvement over the two standard implementations.}
  \label{fig:figure2}
\end{figure}

The SNR metrics for the phantom datasets can be seen in Figure \ref{fig:figure1} and the average SNR across the image, as well as the statistical significance, can be seen in Figure \ref{fig:figure2}. Figure \ref{fig:figure1} shows consistently equal or higher SNR for STRIDE than EDITER A or B in areas of EMI contamination for both the resolution dots and the discs phantom. This is most visible in areas of narrow-band EMI corruption, where EDITER A and EDITER B both have low SNR. Figure \ref{fig:figure2} shows the mean SNR, as well as the statistical significance in the difference between the means. In all 8 cases, STRIDE has significantly better performance than EDITER A, and significantly better performance than EDITER B in five cases. The remaining cases show no significance in their difference.

\begin{figure}[H]
  \centering
  \includegraphics[width=\textwidth]{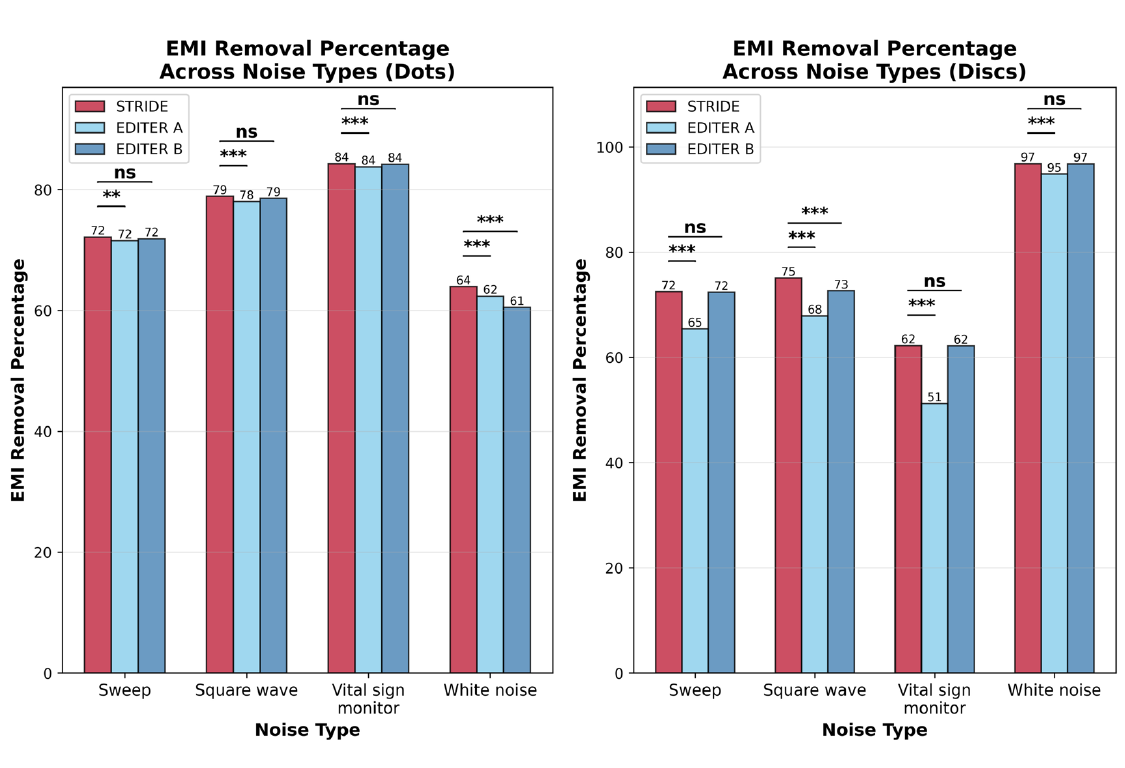}
  \caption{Graphical comparison of the mean EMI removal percentage across non-background voxels for each EMI removal method for\textbf{(left)} the resolution dots and \textbf{(right)} contrast discs in a standard ACR phantom, as well as the statistical significance of the difference in means as determined by a Welch's t-test. $*$ indicates $ p $ <$ 0.05$, $**$ indicates $p < 0.01$, and $***$ indicates $p < 0.001$, and ns indicates no significance. A higher mean EMI removal percentage is better. Overall, there were more cases for which STRIDE significantly outperformed both EDITER A and EDITER B, indicating a statistically significant improvement over the two standard implementations.}
  \label{fig:figure3}
\end{figure}

Similarly, the mean EMI removal percentage can be seen in Figure \ref{fig:figure3}, as well as their statistical significance. In all 8 cases, STRIDE has significantly better performance than EDITER A, and significantly better performance than EDITER B in two cases. In the remaining cases, there is no statistically significant difference.

\begin{figure}[H]
  \centering
  \includegraphics[width=0.9\textwidth]{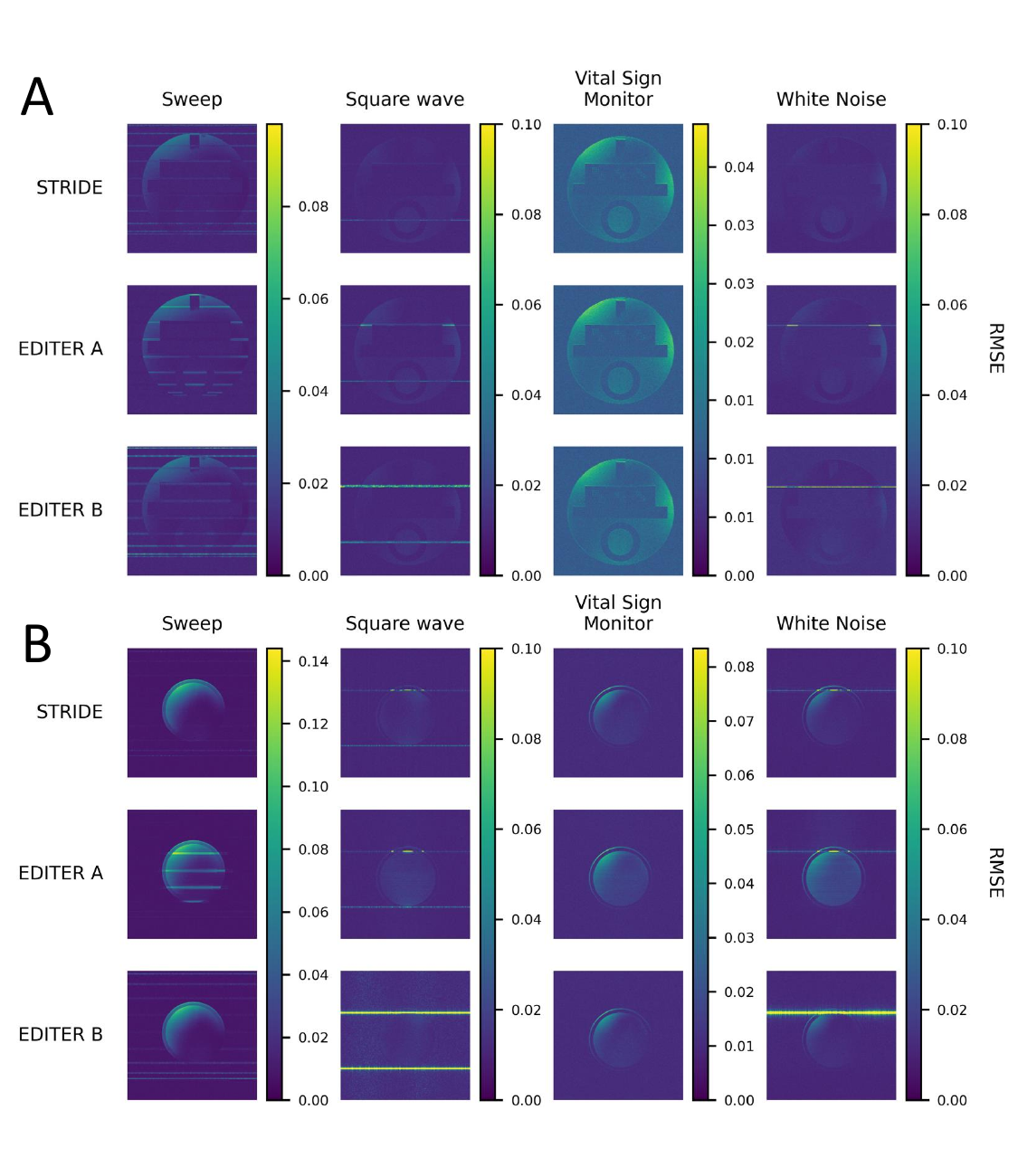}
  \caption{Visual comparison of voxel-wise RMSE for each EMI removal method for \textbf{(A)} resolution dots and \textbf{(B)} contrast discs in a standard ACR phantom. A lower RMSE is better. Overall, STRIDE and the two EDITER implementations have visually similar RMSE. In areas of narrow-band EMI contaimination, however, STRIDE shows visually lower RMSE than either EDITER A or EDITER B. The improvement in RMSE is reflected in Table \ref{tab:RMSEcomparison}, which shows RMSE over the entire imaging volume.}
  \label{fig:figure4}
\end{figure}

RMSE metrics for the phantom datasets can be seen in Figure \ref{fig:figure4}, and the RMSE across the entire 64 image dataset is reported in Table \ref{tab:RMSEcomparison}. Visual inspection of the RMSE images indicates that for both the resolution dots and contrast discs, STRIDE resulted in lower or equal RMSE than EDITER A or EDITER B, specifically along areas of narrow-band EMI. This is reflected in Table \ref{tab:RMSEcomparison}, where STRIDE has a lower RMSE in almost every EMI removal scenario. There is no case in which STRIDE has a higher RMSE than either EDITER A or EDITER B.

\begin{figure}[H]
  \centering
  \includegraphics[width=0.85\textwidth]{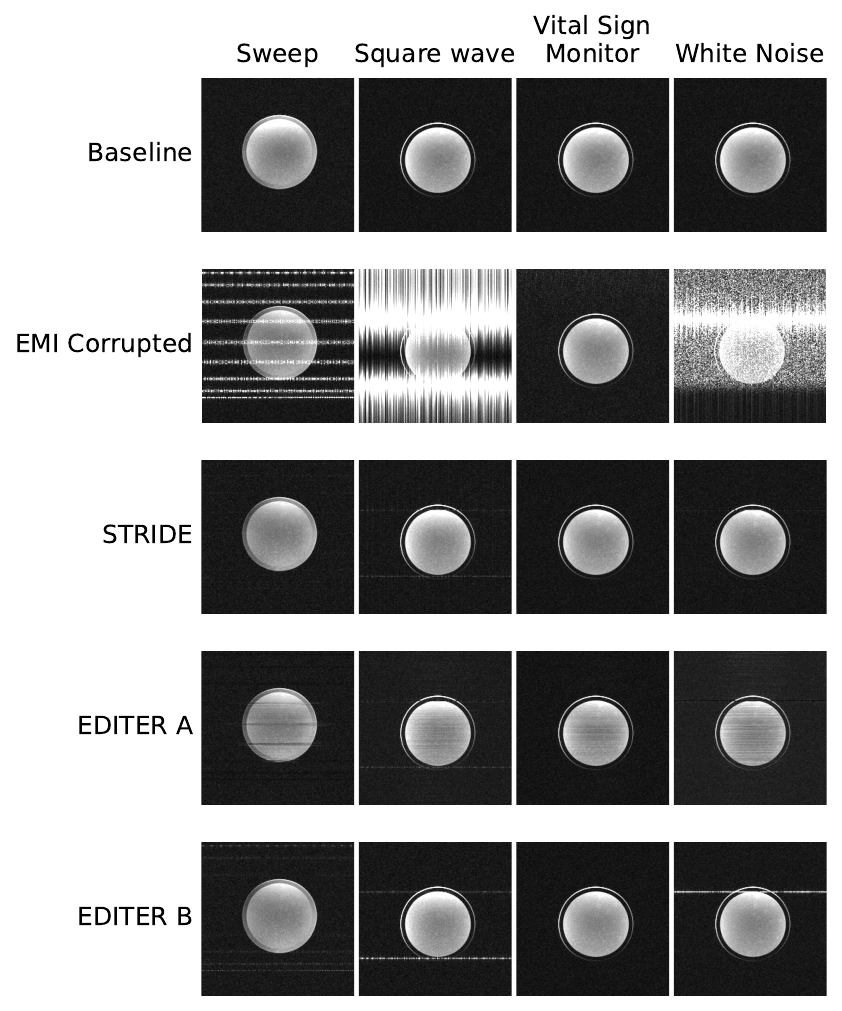}
  \caption{Visual comparison of (first row) baseline images with no EMI corruption, (second row) EMI corrupted images, (third row) EMI corrected images using STRIDE, (fourth row) EMI corrected images using EDITER A, and (fifth row) EMI corrected images using EDITER B for the contrast discs in the small ACR phantom. In every case, EDITER A results in horizontal streaking artifacts across the phantom which are not seen in the STRIDE reconstructions. The performance of STRIDE and EDITER B are similar in the cases of Sweep and Vital Sign Monitor noise. In the cases of Square wave and White noise, STRIDE better removes the EMI than EDITER B.}
  \label{fig:figure5}
\end{figure}

Image reconstruction comparisons for the contrast disc and resolution dots slices can be seen in Figures \ref{fig:figure5} and \ref{fig:figure5} respectively, and for \textit{in-vivo} brain scans in \ref{fig:figure7}. Similar to before, STRIDE had generally better EMI reduction than EDITER A across all EMI scenarios, and better EMI reduction than EDITER B across most scenarios. This can be seen most clearly in scenarios with narrowband noise. With both STRIDE and EDITER B, the resolution dots and contrast discs are generally as visible as in the baseline scans, indicating that these EMI reduction methods didn’t obfuscate any image features or affect overall image resolution or contrast. With EDITER A, however, the dots and discs are less visible due to the streaking artifacts. In the \textit{in-vivo}, STRIDE shows superior EMI removal than either EDITER implementation for narrow-band noise and similar removal for broadband noise.

Figure \ref{fig:figure8} compares reconstructions of square wave EMI corrupted brain images with and without denoising the EMI detection coils. It also shows the standard deviation of the background noise in the highlighted ROI in each case. Here, the reconstruction with denoised EMI sensors vastly outperforms the reconstruction without denoising the EMI sensors both visually and quantitatively with respect to the standard deviation of the noise.

Lastly, Figure \ref{fig:figure9} shows the 15 average image of the brain as denoised by STRIDE, and EDITER B. While the performance is similar, EDITER B leaves visual traces of EMI, whereas STRIDE does not.

\section{Discussion}

We have shown that incorporating total variation into the EMI removal problem formulation results in improved reconstructions when compared to standard EDITER approaches. Overall, STRIDE has significantly superior performance over EDITER in SNR and EMI removal percentage. Even in the single instance when EDITER B outperformed STRIDE in terms of overall SNR, STRIDE had higher local SNR in the EMI corrupted regions. Furthermore, STRIDE always resulted in lower or equal RMSE than EDITER.

This improvement in performance is a direct result of exploiting the smoothness of MR images in their total variation, allowing for more effective EMI removal using a small amount of data per removal, which would typically result in overfitting with conventional methods. The improvement can be seen visually by comparing the EMI corrected images from STRIDE in both phantom datasets, and in the \textit{in-vivo} dataset. In these, STRIDE shows a clear improvement in removing narrow-band EMI over EDITER, and a visually similar performance in removing broadband EMI. The improved narrow band noise removal occurs because the EDITER formulation minimizes MSE. Subtracting the sinusoidal EMI signal from the measured MR signal should eliminate all content at the EMI frequency, creating a dark band in the image at that frequency. However, optimizing for spatial smoothness in the image domain prevents this dark band artifact from appearing.

It is important to note that we used two EMI sensing coils to remove EMI which are fundamentally different from the loop antenna and electrode used in the original EDITER paper, and fewer sensors than used in other EMI removal works \cite{EDITER, Liu2023Electromagnetic, biber2024}. Had we used different or more sensors for detecting EMI, the performance of EDITER B would most likely have increased; however, we believe that the performance of STRIDE would have similarly increased since the addition of the TV matrix is fundamental to the problem and independent of the number or type of EMI detectors. It is also worth noting that while we chose TV as our sparsifying transform in the column direction, other transforms can also be explored (e.g. Wavelets).

Independent of the type of EMI sensor, we have shown the importance of EMI sensor SNR in the EMI removal problem both qualitatively and quantitatively. Our theoretical analysis shows that measurement noise in the EMI sensors will be added to the EMI-free image during EMI removal, which was confirmed in our experiments when comparing reconstructions with and without EMI sensor noise reduction. Although we were able to significantly reduce measurement noise through an SVD-based approach, it is generally not possible since there is no guarantee that the injected EMI will be low rank. Therefore, it is important to either design EMI sensors specifically with measurement noise in mind, or to use many EMI sensors so that measurement noise can be effectively removed through averaging that would occur in transfer function calculation.

The phantom and \textit{in-vivo} datasets used an FA of 90 degrees which is significantly different from the Ernst angle of 43 degrees and 23 degrees for the phantom and brain respectively. A 90 degree flip angle was chosen intentionally to saturate the MR signal, and in-turn cause the EMI to be more pronounced. While this does allow us to better test the EMI removal ability of each algorithm, it resulted in the EMI corrected images to have relatively low SNR.
 
Additionally, we have shown that even under optimal EMI removal conditions and application (15 averages, cleaned EMI sensors,  constant noise source, and coil pre-whitening), the incorporation of total variation into the EMI problem removes EMI artifacts better than the standard EDITER approach.

Since STRIDE operates exclusively in the image domain, it can theoretically be used for any sampling pattern. However, the EMI artifacts created after a non-uniform Fourier transform of measured data to the image domain may be harder to remove in the image domain. In future work, we will test different sampling patterns, and develop computationally efficient algorithms for EMI removal in \textit{k-space} with TV regularization in the image domain.

Although we exploit the inherent smoothness of MR images in the total variation domain, we do not formulate the problem as a typical CS problem. For the sake of computation time and complexity, we relax the standard L1 norm and minimize instead an L2 norm, and the data consistency is implicit. While this does give good results, it may limit reconstruction quality in some cases. It may also not be as compatible with undersampling, where L1 norm based reconstructions are known to outperform L2 based ones.

Lastly, the addition of the total variation matrix in the problem formulation did not increase the computation time per optimization problem relative to the standard EDITER implementations, with all computations happening in roughly one second [2020 M1 MacBook Pro, 8GB RAM].

\section{Conclusion}

We have successfully developed an EMI elimination algorithm that improves over the non-ML popular methods methods. By exploiting inherent MR image smoothness in the total variation domain, our algorithm exhibits overall lower RMSE, higher SNR, and higher EMI artifact removal than standard approaches.

\section{Acknowledgments}
Thank you to Gilbert Thevathasan for constructing the EMI detection coils, and to Tak Fan for helping construct the casing for the EMI detection coils. We would also like to acknowledge the support of CRANIA, and Drs Alexandre Boutet and Taufik Valiante for their support.

\bigskip

\noindent\textbf{Acknowledgments.} Thank you to Gilbert Thevathasan for constructing the EMI detection coils, and to Tak Fan for helping construct the casing for the EMI detection coils. We would also like to acknowledge the support of CRANIA, and Drs. Alexandre Boutet and Taufik Valiante for their support.

\bigskip

\noindent\textbf{Financial disclosure.} The authors acknowledge support from the Ontario Research Fund - Grant: ORF-RE012-009 and INOVAIT - Grant: 2022-4056.

\bigskip

\noindent\textbf{Conflict of interest.} The authors declare no potential conflict of interests.

\bibliographystyle{plain}
\bibliography{references}

\newpage
\begin{table}[H]
\centering
\caption{Tabular comparison of the RMSE across the entire datasets in both the resolution dots and contrast discs slices of the small ACR phantom across all EMI corruption types for the STRIDE, EDITER A, and EDITER B. For both phantom datasets, the ground truth was taken to be the average across all 64 baseline acquisitions. A lower RMSE is better. In most cases, STRIDE outperformed both EDITER A and EDITER B in RMSE. The RMSE for STRIDE was never higher than for EDITER A and EDITER B.}
\begin{tabular}{|l|cc|cc|cc|cc|}
\hline
\multirow{2}{*}{\textbf{Phantom}} & \multicolumn{2}{c|}{\textbf{Sweep}} & \multicolumn{2}{c|}{\textbf{Square wave}} & \multicolumn{2}{c|}{\textbf{WA}} & \multicolumn{2}{c|}{\textbf{White noise}} \\
\cline{2-9}
 & Dots & Discs & Dots & Discs & Dots & Discs & Dots & Discs \\
\hline
STRIDE & 0.13 & 0.016 & 0.12 & 0.012 & 0.12 & 0.012 & 0.13 & 0.013 \\
\hline
EDITER A & 0.14 & 0.020 & 0.12 & 0.014 & 0.13 & 0.013 & 0.14 & 0.016 \\
\hline
EDITER B & 0.15 & 0.018 & 0.14 & 0.048 & 0.12 & 0.012 & 0.16 & 0.076 \\
\hline
\end{tabular}
\label{tab:RMSEcomparison}
\end{table}
\newpage
\newpage

\newpage

\newpage

\newpage

\newpage
\begin{figure}[H]
  \centering
  \includegraphics[width=0.95\textwidth]{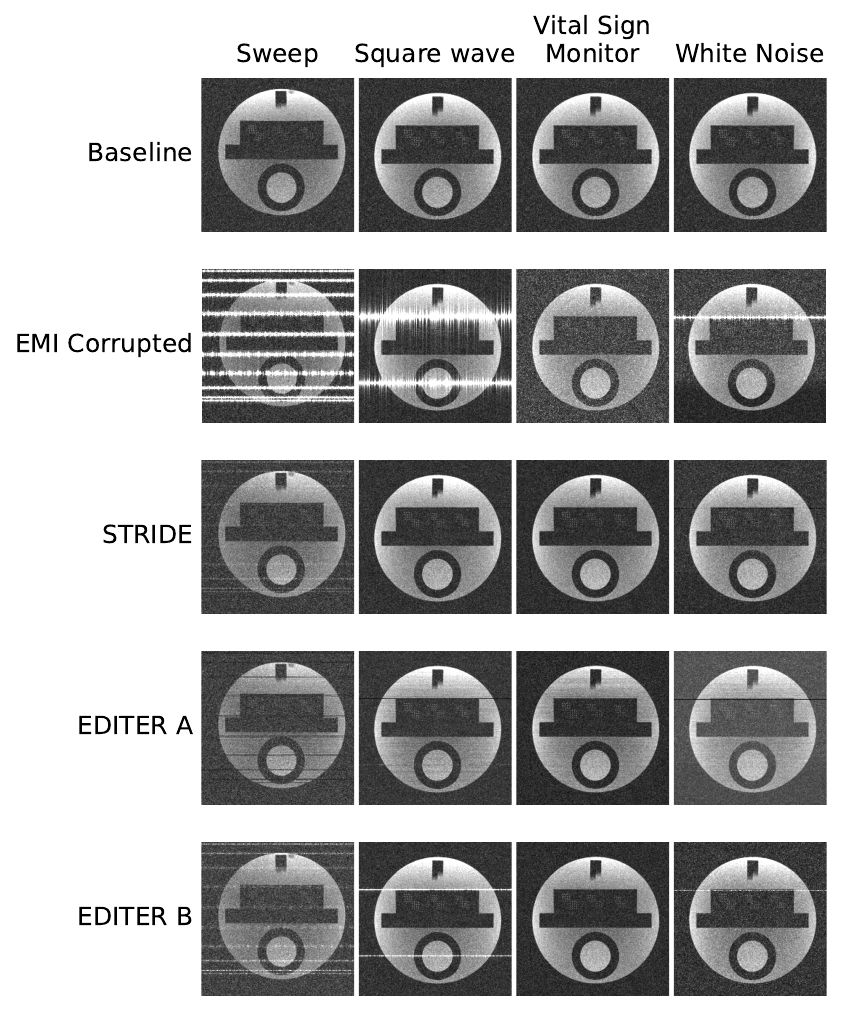}
  \caption{Visual comparison of (first row) baseline images with no EMI corruption, (second row) EMI corrupted images, (third row) EMI corrected images using STRIDE, (fourth row) EMI corrected images using EDITER A, and (fifth row) EMI corrected images using EDITER B for the resolution dots in the small ACR phantom. In every case, EDITER A results in horizontal streaking artifacts across the phantom, as well as dark bands where there was narrow-band EMI corruption, which are not seen in the STRIDE reconstructions. The performance of STRIDE and EDITER B are similar in the cases of Sweep and Vital Sign Monitor noise. In the cases of Square wave and White noise, STRIDE better removes the EMI than EDITER B.}
  \label{fig:figure6}
\end{figure}

\newpage
\begin{figure}[H]
  \centering
  \includegraphics[width=\textwidth]{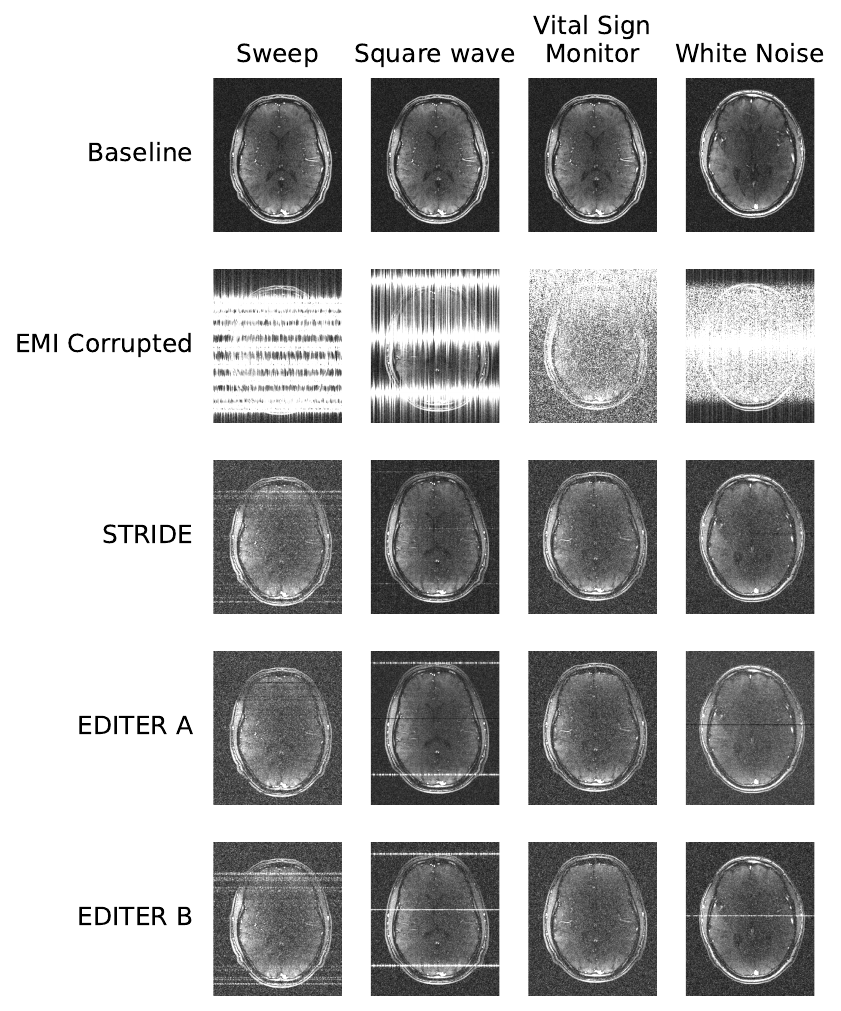}
  \caption{Visual comparison of (first row) baseline images with no EMI corruption, (second row) EMI corrupted images, (third row) EMI corrected images using STRIDE, (fourth row) EMI corrected images using EDITER A, and (fifth row) EMI corrected images using EDITER B for the \textit{in-vivo} brain images. In general, STRIDE better removes narrow-band EMI from the corrupted images than both EDITER A and EDITER B, and has a similar performance to EDITER A and EDITER B in removing broad-band EMI from the corrupted images.}
  \label{fig:figure7}
\end{figure}

\newpage
\begin{figure}[H]
  \centering
  \includegraphics[width=\textwidth]{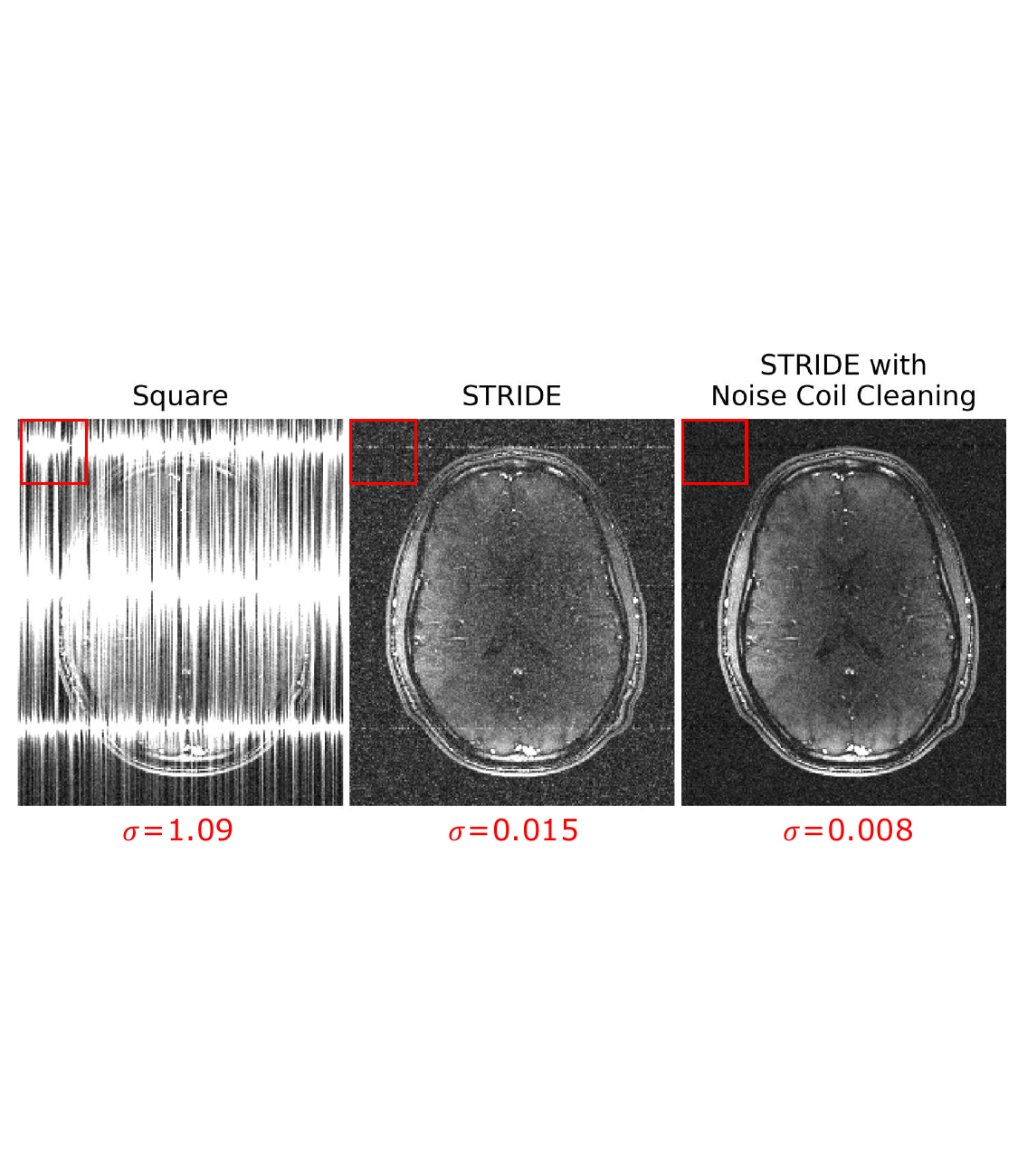}
  \caption{Visual comparison of EMI removal in an \textit{in-vivo} brain image (middle) without any denoising of the two EMI detectors and (right) with SVD denoising of the two EMI detectors compared to (left) the EMI corrupted image. The standard deviation of background noise in the ROI indicated by the red box is reported under the two EMI corrected images. The lower standard deviation in the background noise indicates the improved performance of the EMI removal using the SVD-denoised EMI detectors.}
  \label{fig:figure8}
\end{figure}

\newpage
\begin{figure}[H]
  \centering
  \includegraphics[width=\textwidth]{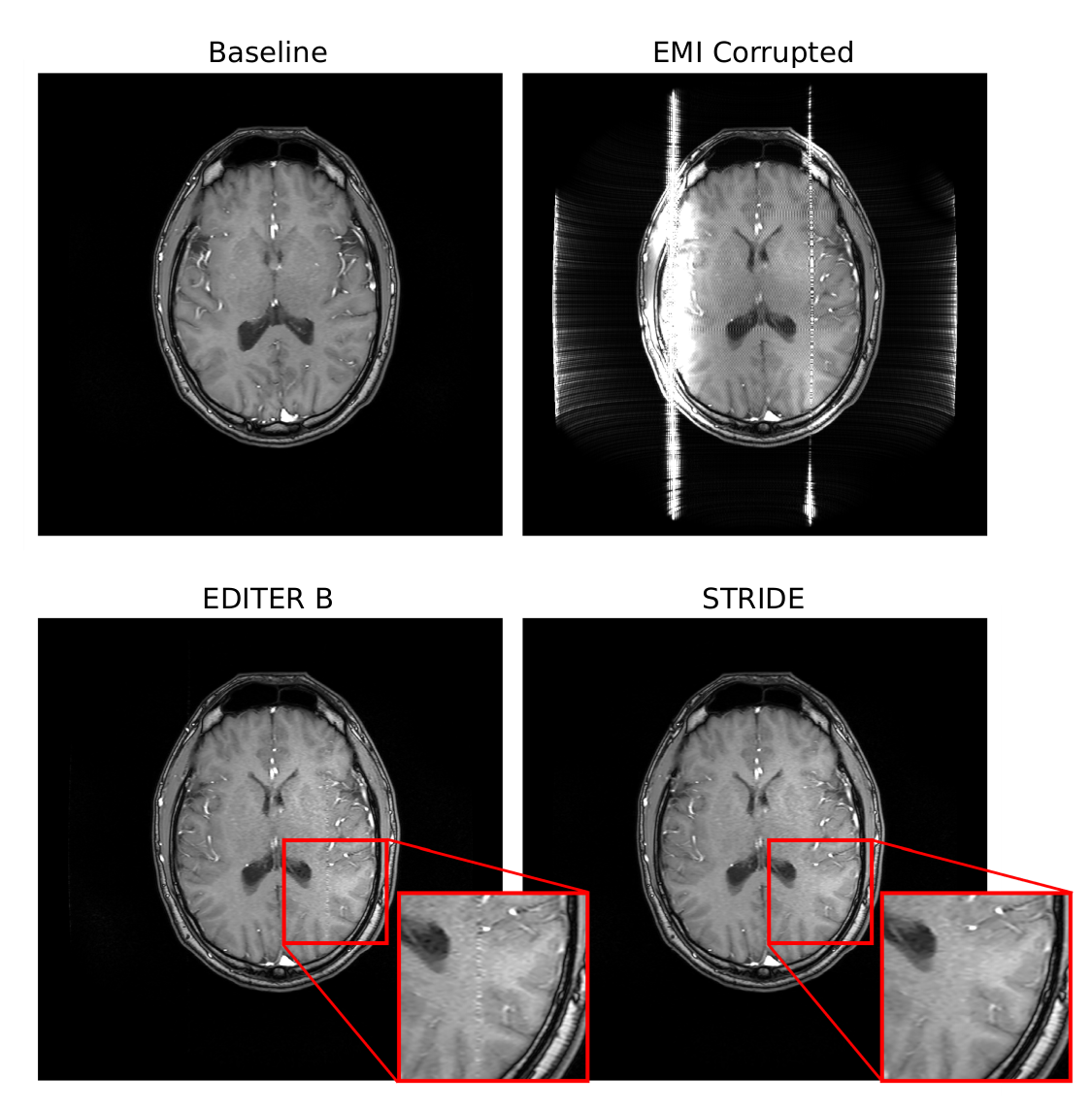}
  \caption{Visual comparison of the (top left) baseline, (top right) EMI corrupted, (bottom right) EMI corrected using STRIDE, and (bottom left) EMI corrected using EDITER B, which is the standard EDITER implementation given by Srinivas et al. \cite{EDITER} for the 15-average \textit{in-vivo} brain dataset. Channel combination was done using the vendor's stock algorithm, which includes prewhitening, shading correcton, and unwarping. The EMI detectors were denoised using SVD-denoising. In the highlighted area, small artifacts from the narrow-band EMI can be seen in the EDITER B EMI corrected image, but not in the STRIDE EMI corrected image.}
  \label{fig:figure9}
\end{figure}

\newpage

\section*{Supporting information}
\begin{figure}[H]
  \setcounter{figure}{0}
  \renewcommand{\thefigure}{S\arabic{figure}\ }
  \centering
  \includegraphics[width=\textwidth]{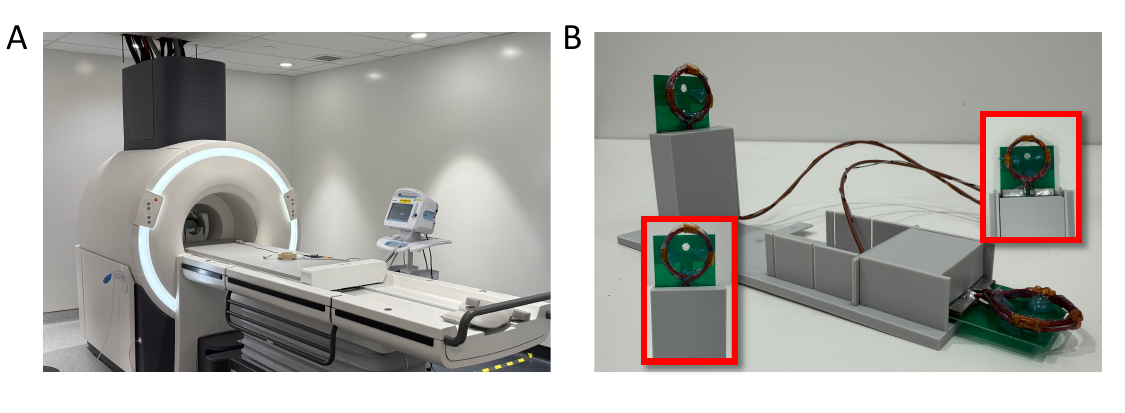}
  \caption{Setup of the WA machine for measuring EMI with the ACR phantom (A), and the EMI sensing coils (B)}
  \label{sup:figure_s1}
\end{figure}

\end{document}